\documentstyle[prl,aps,epsf,epsfig,multicol]{revtex}

\newcommand{\be}{\begin{equation}}
\newcommand{\ee}{\end{equation}}
\newcommand{\bes}{\begin{eqnarray}}
\newcommand{\ees}{\end{eqnarray}}
\newcommand{\bma}{\left( \begin {array}}
\newcommand{\ema}{\end {array} \right)}

\begin{document}

\begin{multicols}{2}

\noindent
{\large \bf Comment on ``Universal Fluctuations in Correlated systems" }

\smallskip

\vspace{-.2cm}

 In \cite{BHP2000} the ``universal" probability density function (PDF)
 of various global measures $x$ from  correlated equilibrium and 
 non-equilibrium physics-inspired models was shown (figure 2 of \cite{BHP2000})
  to be in reasonable agreement with 
 $\Pi (y) = K ( e^{y-e^y})^a$, where $y=b(x-s)$.
     This Bramwell-Holdsworth Pinton (BHP) PDF  with shape parameter $a = \pi/2$, 
was contrasted with the  Fisher-Tippett-Gumbel (FTG) PDF where $a=1$, 
one of 3 possible asymptotes for the maxima  of uncorrelated random variables (r.v.'s) \cite{Colesnotes}. 
In   \cite{BHP2000}  the BHP PDF was also seen to describe the distribution of
maxima $\xi^L_{max}$ from $L$ trials  of  length N vectors $\xi$ of correlated r.v.'s,
   generated by $\xi=\bf{M} \chi$. 
The elements of $\chi$ were i.i.d. exponentially distributed,
 and $\bf{M}$ was  ``an $N \times N$ matrix with random but
fixed elements".
 The  applicability of the BHP PDF  to all the considered  processes in \cite{BHP2000} 
 was inferred to be
   due to ``finite size, {\em strong correlations} (our italics) and self-similarity".
In this Comment we show that this inference is not supported by
the extremal behaviour  of $\xi^L_{max}$: because 1) while a suitable choice of $N$ 
ensures  that $\xi^L_{max}$  falls on the BHP curve, other equally arbitrary
values of $N$  give maxima which don't; and 2) $\xi$ is in fact not strongly correlated,
and the PDF of $\xi^L_{max}$ change little for decorrelated $\xi$.  We thus 
suggest that $a$ differs  from its FTG asymptote of 1 here  largely due to slow
convergence of maxima of the nearly Gaussian $\xi$.
   
In \cite{BHP2000},   ``random" 
   implies ``i.i.d Gaussian, mean 0, variance 1" i.e.   $P(M) = \frac{1}{\sqrt{2 \pi}} 
 e^{-M^2 /2}$ and  ``random but fixed" implies 
 ${\bf M}$ has random elements   held constant between trials. 
 $\xi$ is correlated but the fact that ${\bf M}$ and $\chi$
are statistically independent  means that for the simplest case of $N=2$, $\xi_1$ $(=M_{11}\chi_1 + M_{12}\chi_2)$ has  PDF 
$ P(\xi_1)  =  \int dM_{11} dM_{12} d\chi_1 d\chi_2  \delta (\xi_1 -M_{11}\chi_1 - M_{12}\chi_2)  P(M_{11})P(M_{12})P(\chi_1)P(\chi_2). $
  Taking exponential $P(\chi)= \lambda e^{-\lambda \chi}$ as  in \cite{BHP2000}
  we   generalise to $N \times N$ matrix ${\bf M}$ so
  $  P(\xi) =\frac{1}{2\pi}  \int_{-\infty}^{\infty} I(k)^N e^{ik \xi} dk. $
  Here
$  I(k)^N =  \lambda^N \huge[ \int_{-\infty}^{\infty} \frac{P(M)}{\lambda +i k M} dM  \huge]^N$
 is the generating function of $P(\xi)$.
  We then have
$ I(k) = \sqrt{\frac{\pi}{2}}\frac{\lambda}{k} e^{ \lambda^2/2 k^2} \huge( 1 - erf (\lambda/\sqrt{2}k) \huge).$
   The maxima of $\xi$ are dominated by the tail behaviour of $P(\xi)$, which
  from a small $k$ expansion of $I$ is seen to be Gaussian $\sim (\lambda/\sqrt{N})\exp{-\xi^2 \lambda^2/4N}$.
The maxima of $\xi$ will  thus be in the basin of attraction of the FTG
distribution.  The asymptotic shape parameter $a$ corresponding
to $L,N \rightarrow \infty$  cannot be  changed by
short range dependence \cite{Colesnotes},
but one expects corrections to $a$  for finite $N,L$  because of the
 logarithmically slow convergence of the maxima of Gaussians,
and, sometimes,  a change in  effective degrees of freedom $N$ due
to correlation \cite{Colesnotes}.
 
Figure 1 shows the PDF of $\xi^L_{max}$  for
 $L=10^5$ trials in 2 cases:  $N=40$ and $250$, 
   and for comparison  the PDF  of $\chi^L_{max}=\max(\chi^L)$, when $N=40$, to illustrate the 
      expected    rapid convergence of $\chi^L_{max}$  to the FTG curve. 
  Convergence of $\xi^L_{max}$ however slows with increasing $N$. By $N=40$, convergence of $\xi^L_{max}$ is noticeably
   slower than $\chi^L_{max}$ and the PDF  follows the BHP curve;
while for $N=250$ even slower convergence in $\chi^L_{max}$ gives obvious 
deviation from the BHP form   resembling that seen    in figure 2 of \cite{BHP2000} .
For $N=40$ the autocorrelation
function of $\chi$ (not shown) drops  fast by the first lag. 
Weak (less than $\pm 0.2$) long ranged oscillations {\it are} seen at all lags.
Despite this,  decorrelating $\xi$
by shuffling the $L \times N$ matrix before taking
the maximum of $\xi$ changes the PDF of its maxima in a relatively 
minor way (see Figure 1), confirming that correlation  cannot explain most of the change from the FTG PDF here.
Our result implies that, even though subsequent
results may show that the BHP curve {\it can} result from strong correlation, it {\it need not}.

NWW thanks Mervyn Freeman, Clare Watt, Pat Espy 
and David Stephenson for valuable suggestions.

\begin{figure}[b]
  \begin{center}
    \psfig{file=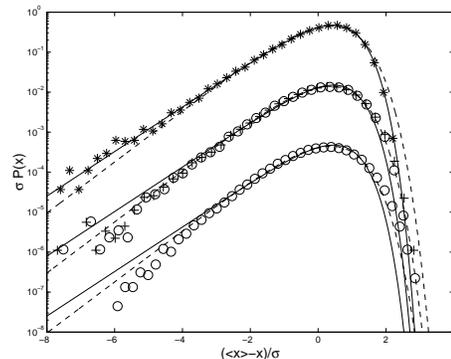,width=6.cm}
    \begin{minipage}{8.5cm}
      \caption{$L=10^5$. Middle plot shows PDF of $\xi^L_{max}$ (circles) compared  with PDF of maxima 
    of decorrelated process (crosses), $N=40$.     
Top plot (stars) is PDF of maxima $\chi^L_{max}$ of exponentially distributed r.v's, $N=40$.  
 Lowest plot (circles) shows $\xi^L_{max}$ in $N=250$ case.
Also shown on each plot are the BHP curve (solid line), and the $a=1$ FTG plot (dashed line).
Displacements of $10^{-1.5}$ and $10^{-3}$ for the middle and lower plots, 
respectively, were applied for clarity. }

\end{minipage}
\end{center}
\label{fig1}
\end{figure}

\vspace{-.8cm}

\noindent
 N. W. Watkins\\
 British Antarctic Survey,  Cambridge, CB3  0ET, U.K.\\
S. C. Chapman and G. Rowlands\\
 University of Warwick, Coventry, CV4 7AL, U.K.\\

\vspace{-.5cm}

Received \today


\end{multicols}

\end{document}